\documentclass[%
 aip,
 amsmath,amssymb,
 reprint,%
]{revtex4-2}

\usepackage{graphicx}
\usepackage{dcolumn}
\usepackage{bm}

\usepackage[utf8]{inputenc}
\usepackage[T1]{fontenc}
\usepackage{mathptmx}
\usepackage{soul}

\newcommand{\imag}{\mathrm{i}}
\newcommand{\dif}{\mathrm{d}}
\newcommand{\rbrac}[1]{\left( #1\right)}
\newcommand{\of}[1]{\left( #1\right)}
\newcommand{\abs}[1]{\left| #1\right|}
\newcommand{\mat}[1]{\underline{#1}}
\newcommand{\bessel}[1]{\mathfrak{J}_{#1}}
\newcommand{\modbessel}[1]{\mathfrak{K}_{#1}}
\newcommand{\rwg}{r_\mathrm{wg}}
\usepackage{import}
\usepackage{graphics}
\usepackage{siunitx}
\usepackage{color}
\usepackage[super]{nth}
\usepackage[english]{babel}

\begin{document}

\preprint{AIP/123-QED}

\title[Existence of a negative next-nearest-neighbor coupling in evanescently coupled dielectric waveguides]{Existence of a negative next-nearest-neighbor coupling in evanescently coupled dielectric waveguides}

\author{J. Schulz}
 \email{schulzj@rhrk.uni-kl.de}
 \affiliation{Physics Department and Research Center OPTIMAS, TU Kaiserslautern, 67663 Kaiserslautern, Germany}
\author{C. Jörg}
\affiliation{Physics Department and Research Center OPTIMAS, TU Kaiserslautern, 67663 Kaiserslautern, Germany}
\affiliation{Department of Physics, The Pennsylvania State University, University Park, Pennsylvania 16802, USA}
\author{G. von Freymann}
\affiliation{Physics Department and Research Center OPTIMAS, TU Kaiserslautern, 67663 Kaiserslautern, Germany}
\affiliation{Fraunhofer Institute for Industrial Mathematics ITWM, 67663 Kaiserslautern, Germany}

\date{\today}

\begin{abstract} 
We experimentally demonstrate that the next-nearest-neighbor(NNN)coupling in an array of waveguides can naturally be negative. To do so, dielectric zig-zag shaped waveguide arrays are fabricated with direct laser writing (DLW). By changing the angle of the zig-zag shape it is possible to tune between positive and negative ratios of nearest and next-nearest-neighbor coupling, which also allows to reduce the impact of the NNN-coupling to zero at the correct respective angle. We describe how the correct higher order coupling constants in tight-binding models can be derived, based on non-orthogonal coupled mode theory. We confirm the existence of negative NNN-couplings experimentally and show the improved accuracy of this refined tight-binding model. The negative NNN-coupling has a noticeable impact especially when higher order coupling terms can no longer be neglected. Our results are also of importance for other discrete systems in which the tight-binding model is often used.

\end{abstract}

\maketitle

\section{\label{Introduction}Introduction}

The tight-binding model is an approximation that is able to reduce the complexity of a system to the point where the dynamics can be described by discrete coupled mode equations. This strong simplification allows to experiment with interesting theoretical models at different physical platforms, for example cold gases~\cite{Haldane_model_Fermions}, electrical circuits~\cite{Topolectrical-circuit}, evanescently coupled dielectric waveguides~\cite{PhD_Tutorial_Szameit} and many more. In waveguides, we can use the tight-binding approximation because the light is mostly confined inside the core of one waveguide. The waveguides nearby are coupled to each other due to the overlap of their evanescent electric fields. Often in these structures, only the couplings to the nearest-neighboring waveguides are considered and those further away are neglected. However, this is valid only in the case of large distances between waveguides, due to the exponential decay of the confined electric field. We take a closer look at the coupling to the next-nearest-neighbors (NNN) in an array of closely spaced waveguides. Thereby we discover that counterintuitively, the NNN-coupling can naturally be negative. Negative couplings to the nearest-neighbor (NN) have been reportedly implemented by detuning the potential~\cite{coupling_between_defects,square-root_topological_insulators,Sign_Control_of_Coupling,fu_topological_2020}, dynamic modulation of the position~\cite{Massless_Dirac_Particles,self-imaging}, or usage of the complex valued field amplitudes of a higher mode~\cite{jorg_artificial_2020}. None of these mechanisms is implemented in our system. It is composed of unmodulated identical single-mode waveguides, but still shows negative coupling. Other previous experimental investigations of the NNN-coupling are either more constraint in their geometry~\cite{photonic_cavities,microwave_artificial_graphene} or have only measured the amplitude of the NNN-coupling~\cite{Second-order_coupling_in_array,Direct_measurement} and assumed it to be naturally positive. The sign of the NNN-coupling, however, is of great importance, since in some systems the NNN-coupling is the defining parameter for the creation of topologically non-trivial phases~\cite{Haldane_model_Fermions,coupled_ring_resonators,generalized_Su-Schrieffer-Heeger}. 

\section{\label{Model}Model}

\begin{figure}[bth]
\def\svgwidth{\columnwidth}
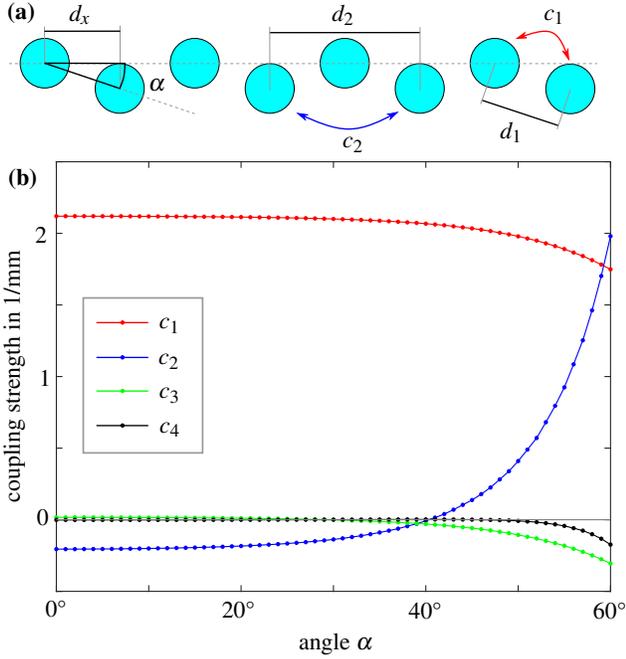
\caption{
(a) Scheme of an angled 1D array of waveguides with an angle $\alpha$, the nearest-neighbor coupling $c_1$ and the next-nearest-neighbor coupling $c_2$. The distance $d_x$ is the size of a unitcell and  $d_1$ and $d_2$ are the distances to the nearest and next-nearest-neighbor, respectively
(b) Coupling coefficients calculated by non-orthogonal coupled mode equations for differently angled arrays of waveguides. The coefficients $c_1$, $c_2$, $c_3$ and $c_4$ are the couplings to the \nth{1}, \nth{2}, \nth{3}  and \nth{4} neighbor, respectively.}
\label{fig:drawing_model}
\end{figure}

We discuss a periodic array of waveguides arranged in a zig-zag shape as shown in Fig.~\ref{fig:drawing_model}(a). By changing the angle $\alpha$, the distance $d_2$ to the NNN can be varied, allowing us to change the NNN-coupling $c_2$ without changing the distance $d_1$ to the NN, thereby keeping the NN-coupling $c_1$ almost constant. Coupling terms to the waveguides further away like to the \nth{3} or \nth{4} neighbor ($c_3$ and $c_4$) are only relevant for large $\alpha$. Calculations based on non-orthogonal coupled mode equations for this arrangement reveal, contrary to the intuition, that the NNN-coupling $c_2$ can be negative, as can be seen in Fig.~\ref{fig:drawing_model}(b). $c_2$ increases with the angle $\alpha$ of the array, as the distance to the NNN decreases. For a straight chain $\alpha=\SI{0}{\degree}$ however, $c_2$ has a negative value, that turns zero at an angle of approximately $\SI{40}{\degree}$. Angles $\alpha>\SI{60}{\degree}$, as in this case the NNNs are closer than the NNs, $d_1<d_2$. The higher order coupling $c_3$ and $c_4$ are close to zero for angles $\alpha<\SI{40}{\degree}$ and are not considered further. The ability to tune the NNN-coupling to zero could be useful also in other experiments if the influence of NNN-coupling is disturbing or unwanted.

The common tight-binding approximation for waveguide structures is based on the assumption that the eigenmode and eigenenergy is mostly determined by the individual waveguide, due to the relatively strong binding by the higher refractive index at the core. The presence of other waveguides nearby is treated as a small perturbation to the original approximation, accounted for by a coupling term that causes the individual eigenenergies to hybridize. This simplification allows us to describe the evolution of the field by the amplitudes $\vec{a}(z)=(...,a_{p-1},a_p(z),a_{p+1}(z),...)^T $ of the eigenmodes in each waveguide $p$ by a system of coupled mode equations

\begin{equation}
\label{eq:CMEq}
-\imag\frac{\partial}{\partial z}\vec{a}\of{z}
=\rbrac{\mat{c}+\mat{\beta}}\vec{a}\of{z}
\end{equation}

where $\mat{\beta}$ is a matrix with the propagation constants on its diagonal. The matrix elements $c_{q,p}$ of the coupling matrix $\mat{c}$ describe the coupling  between two waveguides $q$ and $p$ and are defined by the overlap integral~\cite{PhD_Tutorial_Szameit,photonic_cavities}

\begin{equation}
\label{eq:coupling}
c_{p,q}=k_0\iint{E_q\of{x,y}\Delta n_q\of{x,y} E_p^*\of{x,y} \dif x \: \dif y}
\end{equation}

where $E_q$ and $E_p$ are the transverse field amplitudes for the guided mode for the waveguides $q$ and $p$, respectively and $k_0$ is the wavevector in vacuum. $\Delta n_q$ is the local change in refractive index at waveguide $q$. For a cylindrical waveguide with refractive index $n_\mathrm{core}$ embedded in a material with refractive index $n_\mathrm{clad}$, $\Delta n_q$ is $n_\mathrm{core}-n_\mathrm{clad}$ inside the waveguide and $0$ anywhere else. Note, that for single-mode waveguides Eq.~(\ref{eq:coupling}) only yields positive values, due to the fact that $\Delta n_q$ and the transverse field amplitudes are positive functions. To continue this calculation analytically, for $E_p$ the solution for the field of the eigenmode of a cylindrical waveguide with radius $\rwg$ is used\cite{okamoto_fundamentals_of_optical_wg}

\begin{equation}
\label{eq:eigenmode}
E\of{r}=\frac{1}{\sqrt{N}}
\begin{cases}
\bessel{0}\of{\frac{r}{\rwg}u} 
& r<\rwg \\
\frac{\bessel{0}\of{u}}{{\modbessel{0}}\of{w}} 
\modbessel{0}\of{\frac{r}{\rwg}w}
& r\geq \rwg
\end{cases}
\end{equation}

where $u=\rwg \sqrt{n_\mathrm{core}^2 k_0^2-\beta^2}$, $w=\rwg \sqrt{\beta^2-n_\mathrm{clad}^2 k_0^2}$, $\bessel{n}$ is the Bessel-function of the first kind, $\modbessel{n}$ is the modified Bessel-function of the second kind and $N$ is a factor used to normalize the function. The field and its first derivative have to be continuous at the interface $r=\rwg$, which determines the propagation constant $\beta$ that has to be chosen such that $\bessel{0}\of{u}w\modbessel{1}\of{w}=\modbessel{0}\of{w}u\bessel{1}\of{u}$ is fulfilled. In many cases, where the distances are large between waveguides and higher coupling terms can be neglected, these approximations and equations accurately describe the field along the propagation in waveguide systems.

\subsection{\label{Adjustments}Non-orthogonal tight-binding approximation}

However, under the tight-binding approximation we usually also assume that the transverse fields $E_q$ of the unperturbed eigenmodes form an orthogonal basis, as the overlap is small enough to be neglected. For small distances, when the NNN-coupling can no longer be neglected, also the increasing non-orthogonality has to be taken into account, as stated in Ref.~\cite{non-orthogonal_modes,an_overview}. This is the reason why the coupled mode equations Eq.~(\ref{eq:CMEq}) do not correctly give the coefficient for higher order coupling. To correct this, the non-orthogonality has to be accounted for \cite{okamoto_fundamentals_of_optical_wg}

\begin{equation}
\label{eq:modCMEq}
-\imag\mat{\kappa}\frac{\partial}{\partial z}\vec{a}\of{z}
=\rbrac{\mat{c}+\mat{\kappa}\mat{\beta}}\vec{a}\of{z}
\end{equation}

where $\mat{\kappa}$ is a matrix whose elements $\kappa_{q,p}$ are the overlap integral of the individual eigenmodes

\begin{equation}
\label{eq:kappa}
\kappa_{p,q}=\iint{E_q\of{x,y}E_p^*\of{x,y}\dif x \: \dif y}.
\end{equation}

\begin{figure}[bth]
\def\svgwidth{\columnwidth}
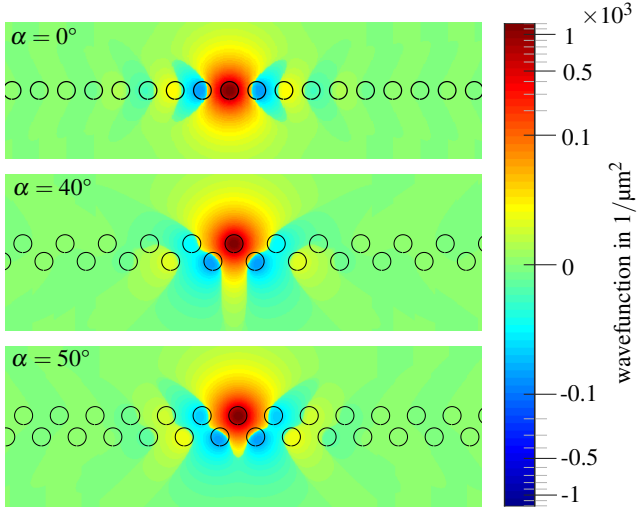
\caption{Orthogonal basis functions for $\vec{b}$ in position space for differently angled arrays.}
\label{fig:basis_function_plot}
\end{figure}

The transformation into an orthogonal basis of localized functions $\vec{b}\of{z}=\mat{T}\vec{a}\of{z}$ can be done with the transformation matrix $\mat{T}$, which is the square root of $\mat{\kappa}=\mat{T}^*\mat{T}$.  After the transformation, the coupled mode equations return to the simpler form of Eq.~(\ref{eq:CMEq})~\cite{base_transformation_non-othorgonal-modes}

\begin{equation}
\label{eq:simCMEq}
-\imag\frac{\partial}{\partial z}\vec{b}\of{z}
=\rbrac{\mat{\tilde{c}}+\mat{\tilde{\beta}}}\vec{b}\of{z}
\end{equation}

where $\mat{\tilde{c}}=\mat{T}^{*-1}\mat{c}\mat{T}^{-1}$ is the corrected coupling matrix and $\mat{\tilde{\beta}}=\mat{T}\mat{\beta}\mat{T}^{-1}$ the corrected propagation constant matrix, in the new basis. By the explicit calculation of the corrected NNN-coupling $\tilde{c}_2=\tilde{c}_{q,q+2}$ and NN-coupling $\tilde{c}_1=\tilde{c}_{q,q+1}$ for our zig-zag shaped array of waveguides, we notice that the ratio $\tilde{c}_2/\tilde{c}_1$ takes positive and negative values dependent on $\alpha$ (see Fig.~\ref{fig:drawing_model}(b)). To stay in the convention that the NN-coupling is positive in dielectric waveguide systems~\cite{coupling_between_defects,square-root_topological_insulators,Sign_Control_of_Coupling,fu_topological_2020}, we will  define the corrected NN-coupling $\tilde{c}_1$ to be positive. It is also worth mentioning that the functions in position space for the basis of $\vec{b}$ exponentially decay over the nearest sites and have an oscillating shape with a period roughly twice the distance $d_1$ to the next neighbor (see Fig.~\ref{fig:basis_function_plot}). The new basis functions share these features  with the Wannier-functions, which form, as the Fourier-transformation of the Bloch-functions, also an orthogonal basis of localized functions.  When these functions for $\vec{b}$ are used for $E_q$ and $E_p$ instead of the unperturbed eigenmodes to calculate the overlap integral Eq.~(\ref{eq:CMEq}), it becomes more intuitive why the NNN-coupling can have a negative sign with respect to  the NN-coupling. Due to the oscillating shape, the overlap also alternates with a period of roughly $2d_1$. This is similar to Ref.~\cite{coupling_between_defects}, where the oscillating shape of the defect mode allows the observation of positive and negative coupling constants dependent on the number of waveguides between the defects.

\subsection{\label{Demonstration}Demonstration of negative NNN-coupling}

To experimentally verify this naturally arising negative NNN-coupling, we fabricate the corresponding waveguide structures, using the direct laser writing principle employed in~\cite{jorg_artificial_2020} (for further details see the Materials and Methods section). We fabricate a sample with three arrays with $\alpha=0\si{\degree}$, $40\si{\degree}$, $50\si{\degree}$ to show the effects for negative, almost zero and positive NNN-coupling, respectively. The negative sign of $c_2$ is not necessarily detectable from the intensity distribution alone, as it mostly affects the phase, but would be noticeable for example in the band structure of the 1D array~\cite{Second-order_coupling_in_array,Long-range_interaction,Nontrivial_Bloch_oscillations}

\begin{equation}
\label{eq:lv2bandstructure}
\beta\of{k_x}
=2c_1 \cos\of{k_x d_x}+2c_2 \cos\of{2k_x d_x}.
\end{equation}

\begin{figure*}[bth]
\def\svgwidth{2\columnwidth}
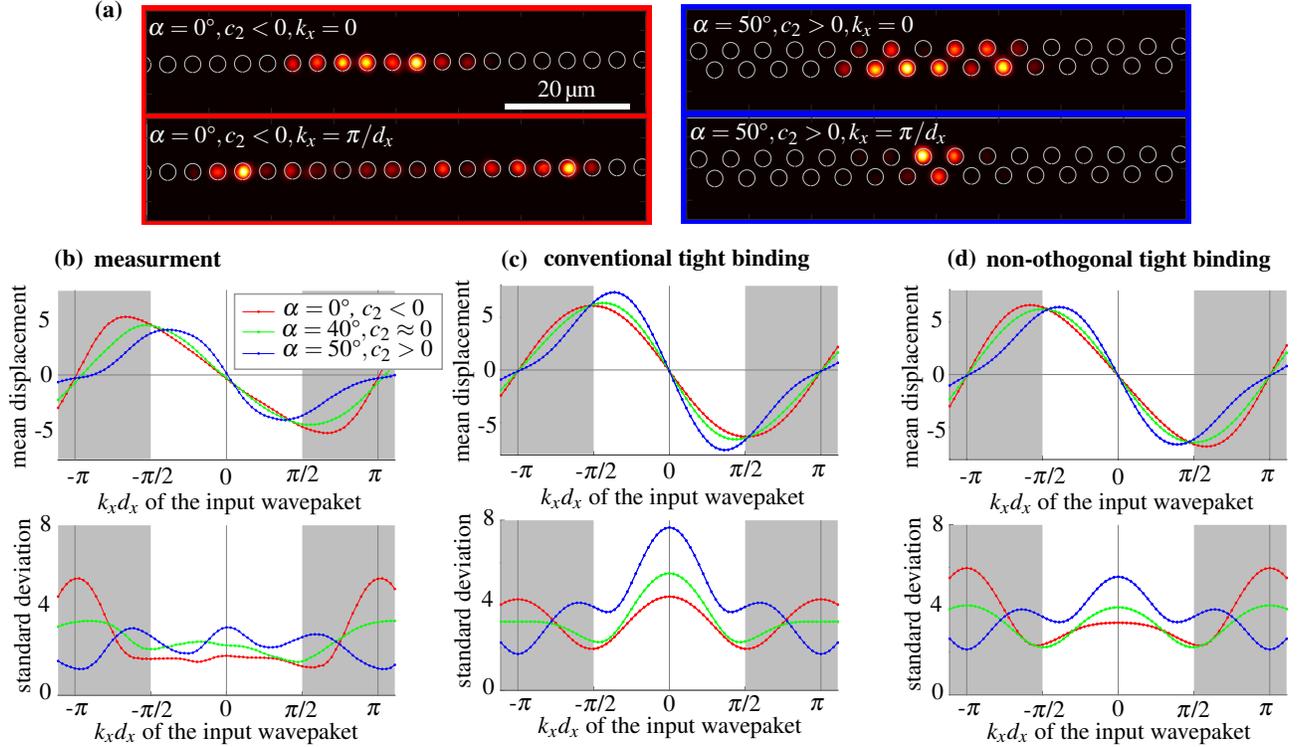
\caption{
(a) Intensity profiles at the output facet of the waveguide structure. The intensity profiles are normalized to their respective maximum value to increase visibility. A greater deviation of a wave packet at the edge of the Brillouin zone at $k_x=\pi/d_x$ compared to one in the center $k_x=0$ points to a negative NNN-coupling. The radius of a waveguides is \SI{1.3}{\micro\metre}, the center-to-center distance to the next waveguide is $d=\SI{4}{\micro\metre}$ and the propagation length is \SI{2}{\milli\metre}. The refractive index contrast of the waveguide to the surrounding material is approximately $\Delta n=0.006$ and the used wavelength is \SI{600}{\nano\metre}.
(b-d) A wave packet is coupled into an array of waveguides with $c_2<0$,$\alpha=\SI{0}{\degree}$ (red); $c_2\approx0$,$\alpha=\SI{40}{\degree}$ (green) and $c_2>0$,$\alpha=\SI{50}{\degree}$ (blue), respectively. The mean displacement and standard deviation of the wave packet with momentum $k_x$ after a propagation distance of \SI{2}{\milli\metre} is 
(b) calculated from the measured intensity profile, 
(c) calculated via the conventional tight binding approximation with matrix $\mat{c}$ and 
(d) calculated by accounting for the non-orthogonality with the matrix $\mat{\tilde{c}}$.
}
\label{fig:measurment}
\end{figure*}

Since we do not have direct access to the band structure in our experiment, we instead measure the group velocity\cite{photonic_parity-time-symmetric_crystals} and the diffraction  of a wave packet with momentum $k_x$. This allows us to gather information about the band structure, as the group velocity and the diffraction are the first and second order derivative of the band structure at $k_x$, respectively. The absolute value of second derivative of Eq.~(\ref{eq:lv2bandstructure}), $\abs{\partial^2 \beta/{\partial k_x^2}}$, determines how much the wave packet spreads along $x$ during propagation through the array in $z$. One sees, that at $k_x=0$ the diffraction is proportional to $\abs{c_1+4c_2}$ while at $k_x=\pi/d_x$ it is proportional to $\abs{c_1-4c_2}$. Therefore, for positive $c_2$, the diffraction of a wave packet at the center of the Brillouin zone is greater than it is at the edge of the Brillouin zone, while for negative values it is the opposite. This effect in the diffraction is clearly visible in the  intensity profiles at the output facet of the structure, shown in Fig.~\ref{fig:measurment}(a), which were captured by a CMOS camera in our experiment. Those images were taken for multiple $k_x$ in the first Brillouin zone and used to calculate the mean displacement and the standard deviation, which quantifies the group velocity and the diffraction, respectively. These results are summarized in Fig.~\ref{fig:measurment}(b). To compare the measurement with the theory, tight binding calculations were performed to predict the intensity profiles at the output. The mean displacement and standard deviation were calculated as shown in Fig.~\ref{fig:measurment}(c,d) where the calculations were based on (c) the standard tight binding approximation where the coupling is always positive and (d) on the corrected coupling constants $\mat{\tilde{c}}$. Note that for negative values of $\tilde{c}_2$ the maximum of the mean displacement shifts to the edge of the Brillouin zone, while it shifts to the center for positive values. This is a feature that only shows in the measurement and the non-orthogonal tight binding model, but is missing in the conventional one. Overall, our measurements qualitatively agree with the corrected coupling constants $\tilde{c}$ and with the negative value of the NNN-coupling.

The negative sign of $c_2$ in a straight array can intuitively be explained by the increasing similarity to a free particle in the limit of short distances between the waveguides. 
As the coupling increases, light is less confined to a single waveguide and the dispersion relation approaches the parabolic shape typical for a free particle $\beta\of{k_x}\propto k_x^2$. Bear in mind that due to the mathematical similarity between Schrödinger and paraxial Helmholtz equation, in our waveguide model system $\beta$ corresponds to the energy in solid state systems. This limit determines what the smallest value of $c_2/c_1$ for identical waveguides can be. The dispersion relation for an array with the coupling constants $c_n$ up to the $N$\textsuperscript{th} neighbor is given by~\cite{Long-range_interaction}

\begin{equation}
\label{eq:lvnbandstructure}
\beta\of{k_x}
=2\sum_{n=1}^N{c_n \cos\of{n k_x d_x}}.
\end{equation}

In the limit of zero spacing between waveguides, the coupling constants $c_n$ in Eq.~(\ref{eq:lvnbandstructure}) have to converge to values of the dispersion relation of a free particle, i.e., a parabolic shape in the first Brillouin zone. By using $\cos \of{m k_x d_x}$ as a basis and the integral over the Brillouin zone as a scalar product in the space of the symmetric and $d_x$-periodic functions, the ratio of the coupling constants in this limit directly follows

\begin{align}
c_m 
&=\int_0^{\pi/d_x}{\beta\of{k_x}\cos\of{m k_x d_x } \dif k_x} \\ 
&\propto \int_0^{\pi/d_x}{k_x^2 \cos \of{m k_x d_x} \dif k_x} 
=2\pi\frac{\rbrac{-1}^m}{m^2}. \nonumber
\end{align}

The smallest possible value of $c_2/c_1$ therefore is -1/4, which indeed has not been undercut by any ratios extracted from our measurements and simulations, further pointing to -1/4 as the lower limit. However, if we are not restricting ourselves to waveguides with equal propagation constants only, this limit can be overcome by detuning the next neighbors, as it has been demonstrated in~\cite{coupling_between_defects,Sign_Control_of_Coupling}. 

Also, this limit of a free particle can explain why at $\alpha=\SI{60}{\degree}$ $c_2$ is bigger than $c_1$ even though $d_1=d_2$ (see Fig.~\ref{fig:drawing_model}(b)). As the light is less confined to a single waveguide the influence of the surrounding refractive index averages out, so that for the light, the waveguide array becomes increasingly similar to a box potential which is elongated in x-direction. The NNN-coupling in our zig-zag array acts (different to the NN-coupling) only along the $x$-direction, which is in the limit of an elongated box potential the direction where the wavefunction is least hindered to spread.

\section{Conclusion}
In conclusion, we explained the counterintuitive negative sign of the NNN-coupling and demonstrated its implications experimentally. We showed that we are able to tune $c_2/c_1$ to zero or even below simply by arranging the waveguides in a zig-zag shape with the corresponding angle. We showed that the simple assumption that the coupling just decreases with distance is not necessarily valid and the non-orthogonality should be accounted for whenever higher-order couplings become relevant. With these results in mind, we can decrease the distance between the waveguides to achieve higher nearest-neighbor coupling, without introducing unwanted effects caused by NNN-coupling. On the other hand, we can look specifically at model systems where these counterintuitive negative coupling values have been dismissed so far. Our results can also be interesting for other discrete systems in which the tight-binding model is often used, as, e.g., in cold gases in optical lattices.

\section{Materials and Methods}

\subsection{Sample fabrication}
The sample is fabricated in IP-Dip using a commercial direct laser writing (DLW) system (Nanoscribe Photonic Professional GT). The refractive index contrast of approximately $\Delta n=0.006$ is achieved by choosing a high laser intensity for the area of the waveguide (LaserPower 75\%) and low intensity for the surrounding area (LaserPower 32\%) close to the polymerization threshold (LaserPower 30\%) at a writing speed of \SI{20}{\milli\metre\per\second}. Thereby, LaserPower 100\% refers to a laser intensity of \SI{55}{\milli\watt} before the 63$\times$ focusing objective of the DLW system. The radius of a waveguide is \SI{1.3}{\micro\metre}, the center-to-center distance to the next waveguide is $d=\SI{4}{\micro\metre}$ and the propagation length is \SI{2}{\milli\metre}. After the writing process, excess photoresist on top of the sample, that would cause distortions at the measurement, is removed by dipping the top into PGMEA for a minute. As a last step, the sample is exposed to UV light from an Omnicure S2000 with 95\% iris opening for 60s to cure excess photoresist.

\subsection{Measurement}
For the measurement, laser light at a wavelength of \SI{600}{\nano\metre} from a white light laser (NKT photonics) and a VARIA filter box is used. The light is linearly polarized, expanded and sent onto a spatial light modulator (SLM). Afterwards, all light besides the first diffraction order of the blazed grating from the SLM is blocked. With a 20$\times$ objective (NA=0.4), the Fourier transformed hologram from the SLM is imaged onto the input facet of the waveguide sample. The intensity and phase profile on the sample’s input facet can be tuned by adjusting the hologram. To couple a wave packet with momentum $k_x$ into the waveguide array, light is focused into seven neighboring waveguides with a Gaussian-shaped intensity-envelope and a phase difference between each focus corresponding to the $k_x$ we want to measure. The intensity of the light from the output facet of the waveguide structure is finally imaged via a second 20$\times$ objective onto a CMOS camera (Thorlabs DDC1545M).

\begin{acknowledgments}

C.J.\ gratefully acknowledges funding from the Alexander von Humboldt Foundation within the Feodor-Lynen Fellowship program. G.v.F.\ acknowledges funding by the Deutsche Forschungsgemeinschaft through CRC/Transregio 185 OSCAR (project No.\ 277625399).

\end{acknowledgments}

\section{Data availability}
The data that support the findings of this study are available from the corresponding authors upon reasonable request

\bibliography{mybib}

\end{document}